\documentclass[journal=nalefd,manuscript=letter]{achemso}

\usepackage[version=3]{mhchem}
\usepackage{graphicx}

\author{Davide Staedler} \affiliation{Institute of Chemical Sciences and Engineering, EPFL, CH-1015, Lausanne, Switzerland}\altaffiliation{Contributed equally to this work.}

\author{Thibaud Magouroux} \affiliation{GAP-Biophotonics, Universit\'e de Gen\`eve, 22 chemin de Pinchat, CH-1211 Gen\`eve 4, Switzerland}\altaffiliation{Contributed equally to this work.}

\author{Sol\`ene Passemard}  \affiliation{Institute of Chemical Sciences and Engineering, EPFL, CH-1015, Lausanne, Switzerland}

\author{Sebastian Schwung} \affiliation{FEE Gmbh, Struthstrasse 2, 55743 Idar-Oberstein, Germany}

\author{Marc Dubled} \affiliation{SYMME, Universit\'e de Savoie, BP 80439, 74944, Annecy Le Vieux Cedex, France}

\author{Guillaume St\'ephane Schneiter} \affiliation{Institute of Chemical Sciences and Engineering, EPFL, CH-1015, Lausanne, Switzerland}

\author{Daniel Rytz} \affiliation{FEE Gmbh, Struthstrasse 2, 55743 Idar-Oberstein, Germany}

\author{Sandrine Gerber-Lemaire} \affiliation{Institute of Chemical Sciences and Engineering, EPFL, CH-1015, Lausanne, Switzerland}

\author{Luigi Bonacina} \affiliation{GAP-Biophotonics, Universit\'e de Gen\`eve, 22 chemin de Pinchat, CH-1211 Gen\`eve 4, Switzerland}\phone{+41 22 3790508} \email{luigi.bonacina@unige.ch}

\author{Jean-Pierre Wolf} \affiliation{GAP-Biophotonics, Universit\'e de Gen\`eve, 22 chemin de Pinchat, CH-1211 Gen\`eve 4, Switzerland}

\title{Frequency Doubling Nanocrystals for Cancer Theranostics }

\begin{document}

\begin{abstract}

A  novel bio-photonics approach based on the nonlinear optical process of second harmonic generation by non-centrosymmetric nanoparticles is presented and demonstrated on malignant human cell lines. The proposed method allows to directly interact with DNA in absence of photosensitizing molecules, to enable independent imaging and therapeutic modalities switching between the two modes of operation by simply tuning the excitation laser wavelength, and to avoid any risk of spontaneous activation by any natural or artificial light source.
\end{abstract}

We demonstrate here a  novel diagnostic and therapeutic (\textit{theranostic}) protocol  based on the nonlinear optical process of non phase-matched second harmonic (SH) generation by non-centrosymmetric nanoparticles, referred to in the following as harmonic nanoparticles (HNPs).\cite{Bonacina, Dempsey} To date, the capability of these recently introduced nanometric probes of doubling \textit{any} incoming frequency has not been  employed for therapeutic use, although it presents several straightforward advantages, including i) the possibility to directly interact with DNA of malignant cells in absence of photosensitizing molecules, ii) fully independent access to imaging and therapeutic modalities, and iii) complete absence of risk of spontaneous activation by natural or artificial light sources other than pulsed femtosecond lasers. Given the unconstrained tunability of the HNPs nonlinear conversion process, this approach can be extended to selectively photo-activate molecules at the surface or in the vicinity of HNPs to further diversify the prospective therapeutic action.\cite{Zhao} Here we show that by tuning the frequency of ultrashort laser pulses from infrared (IR) to visible (both harmless), SH generation leads respectively to diagnostics (imaging) and therapy (phototoxicity). Specifically, we report \textit{in situ} generation of deep ultraviolet (DUV) radiation (270 nm) in human-derived lung cancer cells treated with bismuth ferrite (\ce{BiFeO3}, BFO) HNPs upon pulsed laser irradiation in the visible spectrum, at 540 nm. We observe and quantify the appearance of double-strand breaks (DSBs) in the DNA and cell apoptosis, in the area of the laser beam. We show that DNA damages are dependent on irradiation-time, laser intensity, and NP concentration. We observe that apoptosis and genotoxic effects are only observed when visible light excitation is employed, being completely absent when IR excitation is used for imaging.

\begin{figure}
\begin{center}
\includegraphics[width=8 cm]{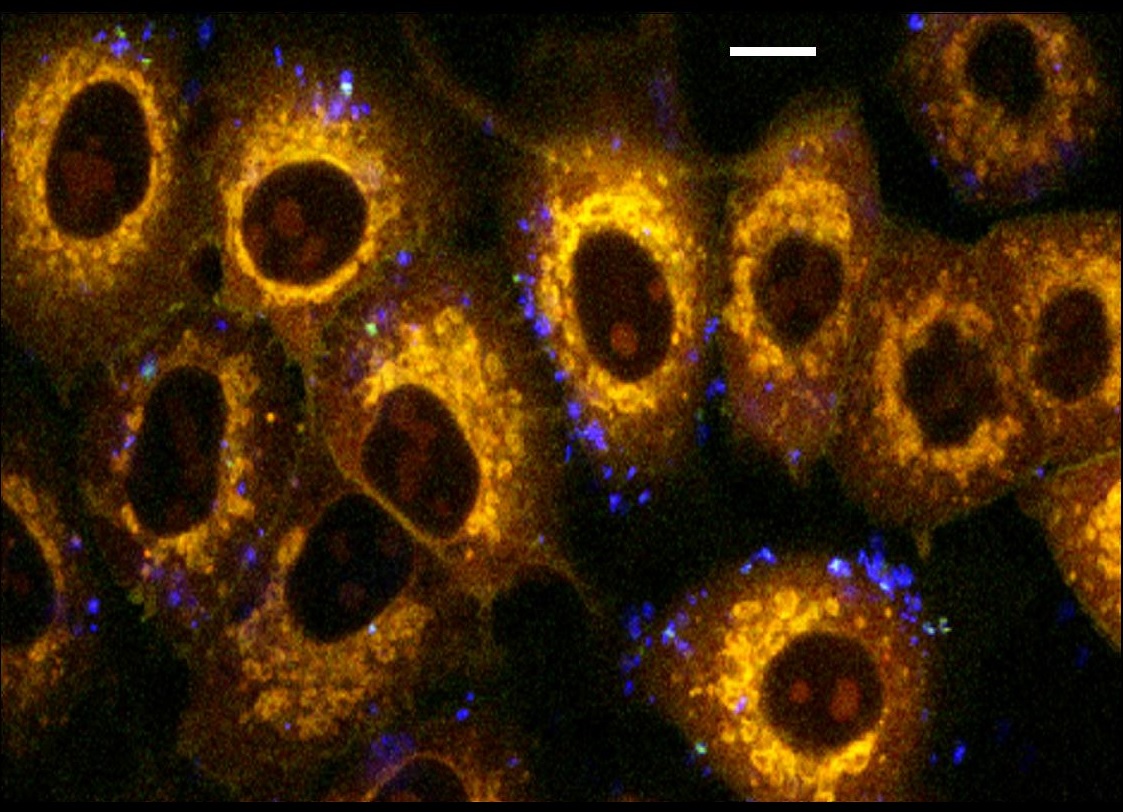}\caption{\textbf{Multiphoton imaging of a HNPs treated sample.} Lung-derived A549 cancer cells exposed for 5 h to 50 $\mu$g/mL BFO HNPs. Yellow: two photon excited fluorescence from cell membrane dye FM1-43FX. Blue: SH signal from HNPs. Scale bar:10 $\mu$m.}
\label{Fig:1}
\end{center}
\end{figure}

\begin{figure}
\begin{center}
\includegraphics[width=7 cm]{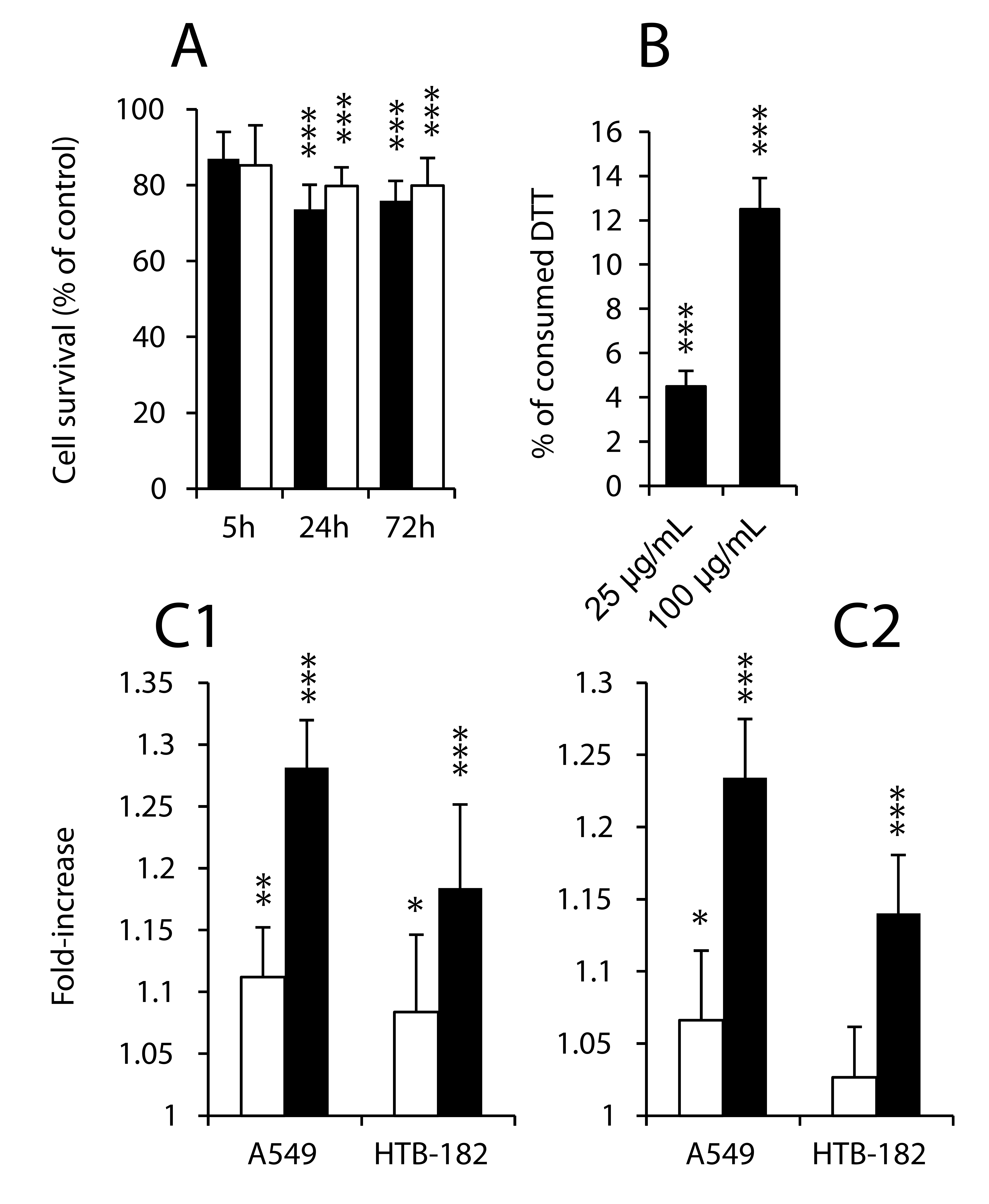}\caption{\textbf{Cytotoxic and oxidative effects of BFO HNPs. The nanomaterial shows good bio-compatibility with high survival rate after long-term incubation.} \textbf{A:} cell survival by MTT for different exposure times of lung-derived A549 (black bars) and HTB-182 (white bars) cancer cell lines to 100 $\mu$g/mL BFO. \textbf{B:} effect of BFO at 25 and 100 $\mu$g/mL on DTT content in an abiotic environment after 1 h of incubation. \textbf{C1} and \textbf{C2:} ROS production after 24 h exposure to 25 (white bars) or 100 (black bars) $\mu$g/mL HNPs by DHE (\textbf{C1}) and DCFH-DA (\textbf{C2}) assays. Results statistically compared to untreated cells.
}
\label{Fig:2}
\end{center}
\end{figure}

HNPs, a family of NPs specifically conceived for multi-photon imaging, were  introduced in 2005 for complementing fluorescence imaging labels.\cite{Bonacina, Nakayama,  PantazisPNAS} Although comparatively less bright than quantum dots, HNPs possess a series of advantageous optical properties, including complete absence of bleaching and blinking\cite{LeXuan, Bonacina}, spectrally narrow emission bands, fully coherent response,\cite{Baumner, Hsieh1, Hsieh2} ,and UV to IR excitation wavelength tunability.\cite{Extermann, Staedler} These unique characteristics have been recently exploited in demanding bio-imaging applications\cite{Culic} including regenerative research.\cite{Magouroux} The possibility of working with long wavelengths presents clear advantages in terms of tissue penetration, as in the IR spectral region, imaging  depth is strongly increased by reduced absorption (provided that water absorption is avoided) and weak scattering (preventing degradation of spatial and temporal laser profiles).\cite{Zipfel}

As an example of HNPs based imaging, Fig. \ref{Fig:1} displays lung-derived A549 cancer cells stained with FM1-43FX cell membrane dye exposed for 5 h to BFO HNPs at 50 $\mu$g/mL. The image was acquired upon near IR excitation at 790 nm, the two-photon excited fluorescence from the dye are shown in yellow, while the intense blue spots correspond to SH radiation emitted by HNPs. The latter have the tendency to remain attached to cell membranes without being internalized due to their relatively large size. As for other nanobiotechnological approaches, selective binding of NPs to specific cell membrane receptors would rely on the presence of targeting molecules at their surface,\cite{Culic} a strategy that was not implemented in this exploratory study.

Given the novelty of the nanomaterial employed in this study, prior to the assessment of photo-therapeutic modality, BFO HNPs were characterized and screened for biocompatibility in terms of cytotoxicity and oxidative effect. The cytotoxic effect of 100 $\mu$g/mL BFO HNPs was assessed after 5, 24 and 72 h exposure (Fig. \ref{Fig:2}A) on two lung-derived cell lines (A549, HTB-182). BFO cytotoxicity was found acceptable in both samples as HNPs did not cause any detectable effect on cell survival after 5 h exposure, and after 24 h and 72 h  cell viability  remains remarkably high (>75\%), comparable to that observed with HNPs composed of other nanomaterials previously screened.\cite{Staedler} To quantify the oxidative stress induced by BFO HNPs, the catalytic activity of the NPs was first measured in a cell free environment by dithiothreitol (DTT) assay.\cite{Cho} BFO HNPs show a dose-dependent consumption of DTT after 1 h incubation (Fig. \ref{Fig:2}B), suggesting that they can exert catalytic production of superoxide.  The production of  reactive oxygen species (ROS) by BFO HNPs in cell cultures was assessed using two fluorescence assays: dihydroethidium (DHE) and carboxydichlorodihydrofluorescein diacetate (DCFH-DA).\cite{Frick, AshaRani} We could detect a dose-dependent increase of ROS, more pronounced in A549 cells than in HTB-182 (Fig. \ref{Fig:2}C1 and C2), which remains  however low compared to that induced by other metal-based NPs \cite{Staedler, Barzilai, Frick, AshaRani}. Overall, the result of this thorough screening indicates a good biocompatibility of this nanomaterial, tested for the very first time for biological applications, and sets the ground for the light-triggered HNPs-cells interaction described in the following.

\begin{figure}
\begin{center}
\includegraphics[width=8 cm]{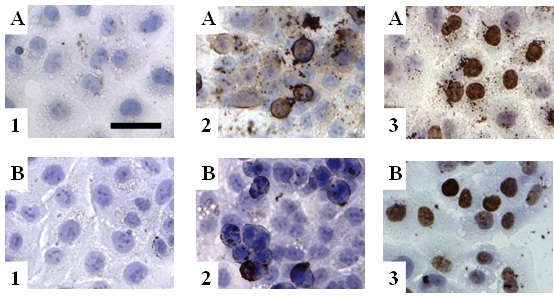}\caption{\textbf{ Immunohistochemistry images demonstrating the effect of visible irradiation on malignant cell lines for apoptosis (cPARP) and DNA repair ($\gamma$H2AX).}
Human lung-derived A549 (\textbf{A}) and HTB-182 (\textbf{B}) cancer cell lines untreated (\textbf{1}) or exposed to 100 $\mu$g/mL BFO HNPs (\textbf{2} and \textbf{3}), incubated 24 h, and irradiated for 120 s. Expression of cPARP (\textbf{2}) or $\gamma$H2AX (\textbf{3}) observed by IHC after further 24 h or 30 min of incubation, respectively. Positive cells are in brown, nuclei in blue, and HNPs  aggregates appear as small brown spots. Scale bar: 50 $\mu$m}
\label{Fig:3}
\end{center}
\end{figure}

DNA absorption is particularly efficient in the deep UV (DUV, <300 nm), as all DNA bases possess bands peaking around 260 nm with negligible intensity from 310 nm on. Irradiation of cell cultures at this wavelength results into DSBs and evokes a complex network of molecular responses, eventually resulting in DNA repair and/or cell apoptosis.\cite{Hanasoge, Yuan, Stergiou} Histone variant H2AX is a key component of the early stage response to DNA damages, as upon UV exposure it is phosphorylated at its carboxyl terminus to form  $\gamma$H2AX at the DSBs sites.\cite{Hanasoge, Yuan, Lu, Shih} After the first appearance of UV-induced DNA-damages, cells first activate DNA-repair mechanisms and then apoptosis occurs to eliminate potentially hazardous cells. UV-dependent apoptosis is caused by the activation of caspase-3 and subsequently cleavage of the poly (ADP-ribose) polymerase (PARP), resulting in a cleaved-form (cPARP) with a mass of 89 kDA.\cite{Shih, Takasawa, Tuchinda} For the irradiation experiment, cells were plated in 35 mm Petri dishes with glass bottom for 48 h, then medium was replaced and cells were incubated for 24 h with BFO HNPs (25 or 100 $\mu$g/mL, 2 mg/mL stock solution in water) or a negative control containing the vehicle (distilled water).  The sample was exposed for 30, 60, or 120 s to ultrashort (30 fs) pulses of visible light generated by a noncolliner OPA (15 mW average power, 1 KHz repetition rate) with a laser spot size of 170 $\mu$m diameter. During  irradiation, cells were kept in a microscope incubator. After light treatment,  cells were incubated for 30 min ($\gamma$H2AX assay) or 24 h (cPARP detection)\cite{Lu, Lee, Narayanapillai} and then fixed with 3\% formaldehyde in PBS. Frequency doubling of femtosecond pulses of visible light (540 nm) by HNPs attached to cell membranes generates DUV photons in the close vicinity of cell nuclei, optimally placed for direct photo-interaction (see Fig. \ref{Fig:1}). The biological effects of such laser irradiation is reported in the immunohistochemistry (IHC) images of Fig. \ref{Fig:3}. The two image rows are associated to the two  human malignant cell lines already tested for cytotoxicity, A549 and HTB-182. The control samples (A1, B1) show no expression for both reporters, confirming that they are not present under physiological conditions,   while the clear effect (positive cells in brown) visible in panels 2 and 3 for treated cells indicates that a strong interaction upon irradiation takes places, showing the DNA-repairing enzyme $\gamma$H2AX expression well localized within the nuclei and the cPARP reporter of cell apoptosis in the cytoplasm of the damaged cells.

\begin{figure}
\begin{center}
\includegraphics[width=6 cm]{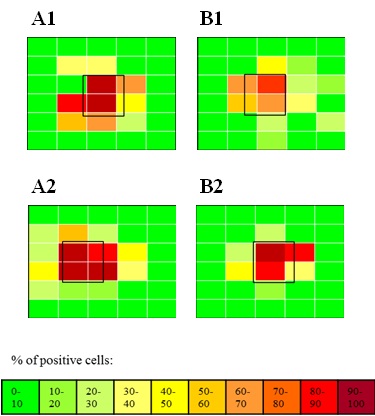}\caption{\textbf{Spatial localization of the expression of  cPARP and $\gamma$H2AX reporters.} Analysis of one representative IHC  image (480 x 650 $\mu$m) for the treatments in Fig. \ref{Fig:3}. For each rectangular (80 x 130 $\mu$m) sub-region the \% ratio of cells positive for the expression of cPARP (column 1) and $\gamma$H2AX (column 2) is expressed according to the color-scale. Black empty square: laser spot (150 x 150 $\mu$m).
}
\label{Fig:5}
\end{center}
\end{figure}

The spatial localization of the IHC expression of the two reporters for DNA repair and cell apoptosis is given in Fig. \ref{Fig:5}. The laser focal spot (empty black square) is superimposed to a spatially resolved pattern of $80\times 130 \mu$m rectangles indicating in false colors the \% of positive cells to  cPARP (A1, A2) and $\gamma$H2AX (B1, B2) for the two cell lines. One can appreciate how the biological effect of visible irradiation perfectly co-localizes with the laser spot (>80 \% positive cells) and rapidly decreases outside the focal region  to negligible values.

\begin{figure}
\begin{center}
\includegraphics[width=7 cm]{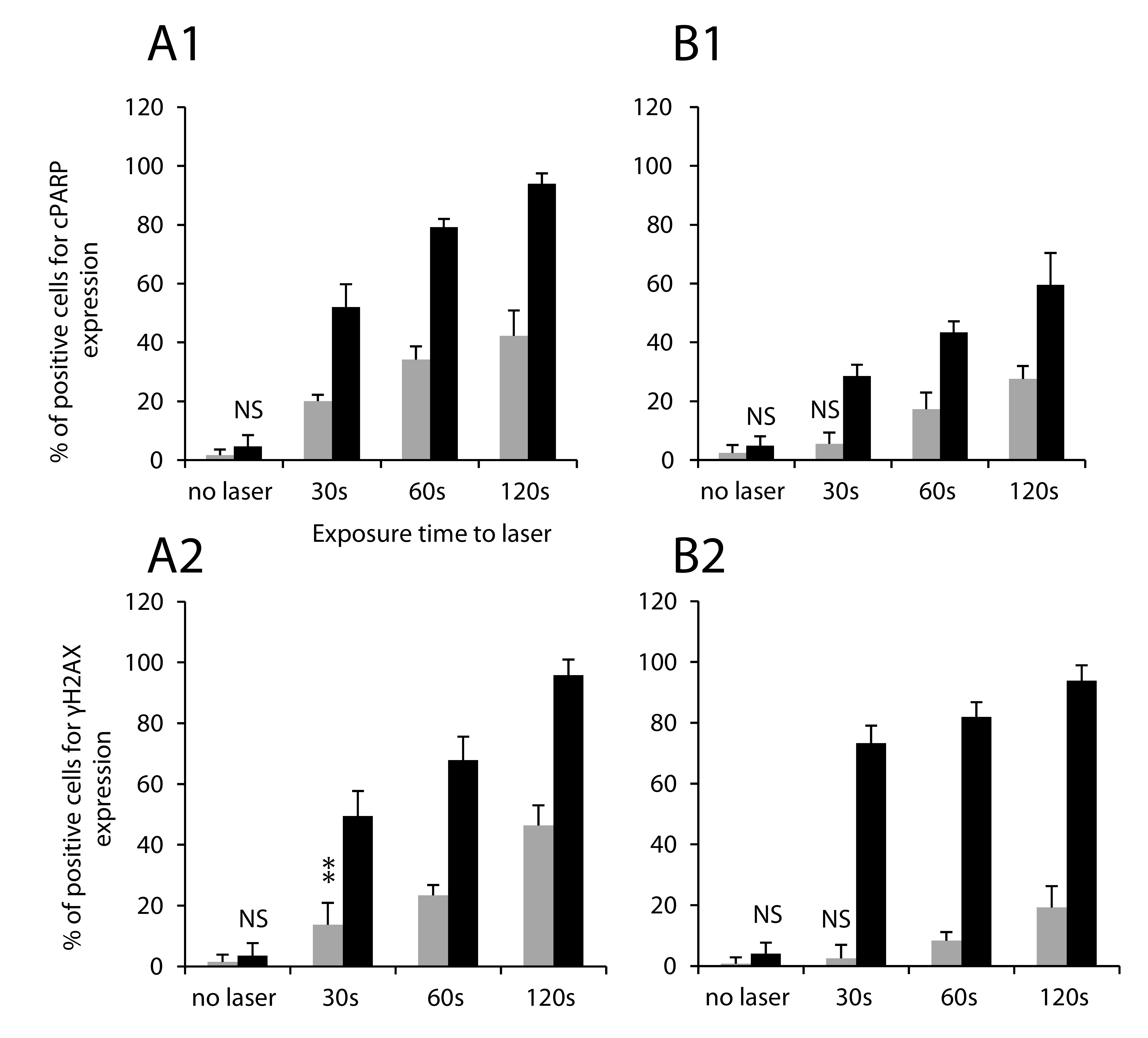}\caption{\textbf{Apoptosis and DNA damage upon visible light exposure.} Human lung-derived A549 (A) and HTB-182 (B) cancer cell lines exposed to vehicle (grey bars) or 100 $\mu$g/mL BFO HNPs (black bars), incubated 24 h and irradiated for 30, 60, or 120s. Expression of cPARP (upper row) or $\gamma$H2AX (lower row) was observed by IHC after further 24 h or 30 min of incubation, respectively. Results are expressed as \% ratio of positive cells. All comparisons to control or between cells exposed to laser with and without BFO HNPs are significant (p<0.001) if not otherwise specified.}
\label{Fig:4}
\end{center}
\end{figure}

The quantitative assessment of the effects of \textit{in situ} DUV generation is reported in Fig. \ref{Fig:4}. In the histograms, the number of IHC-positive cells is expressed as \% ratio of total cells in the area of the laser-spot. Firstly, one can observe that BFO HNPs at 100 $\mu$g/mL without laser irradiation do not cause any increase of cPARP and $\gamma$H2AX expression, ensuring that  oxidative effects of BFO alone do not interfere with irradiation assays. Upon laser-exposure, the ratio of cells positive for cPARP and $\gamma$H2AX clearly increases in an exposure time-dependent manner. In A549 the expression of the two proteins is comparable, whereas HTB-182 express systematically more $\gamma$H2AX. Such a stronger enzymatic activity seems correlated with  higher cell viability: the maximal expression of cPARP (apoptosis) is around 60\% while in A549 it reaches almost 100\%.

The major decrease in cells viability observed upon irradiation of HNPs-treated samples, together with the high spatial localization of the biological effects, makes HNPs-based approaches amenable for developing therapeutic (photo-dynamic) protocols. With this goal in mind, we performed an additional verification, essential to ensure the possibility of independently addressing imaging and cell irradiation modalities, ensuring that the definition of the zone to be treated (which might rely on a specific NPs surface functionalization) can be preliminary safely performed without any risk of unwanted activation.  Cells exposed to HNPs were irradiated for 5 min with laser set at 790 nm (SHG at 395 nm, outside  the DNA bases absorption band) with intensity parameters equal to those of the protocol described above for visible irradiation. In this case, we did not remark any interference on cells metabolism. As reported in previous works, imaging is not limited to near infrared wavelengths but it can be performed even above 1.5 $\mu$m, with clear advantages in terms of  imaging depth (thanks to decreased scattering and absorption) and long-term photo-stability.\cite{Bonacina, Extermann}

\begin{figure}
\begin{center}
\includegraphics[width=7 cm]{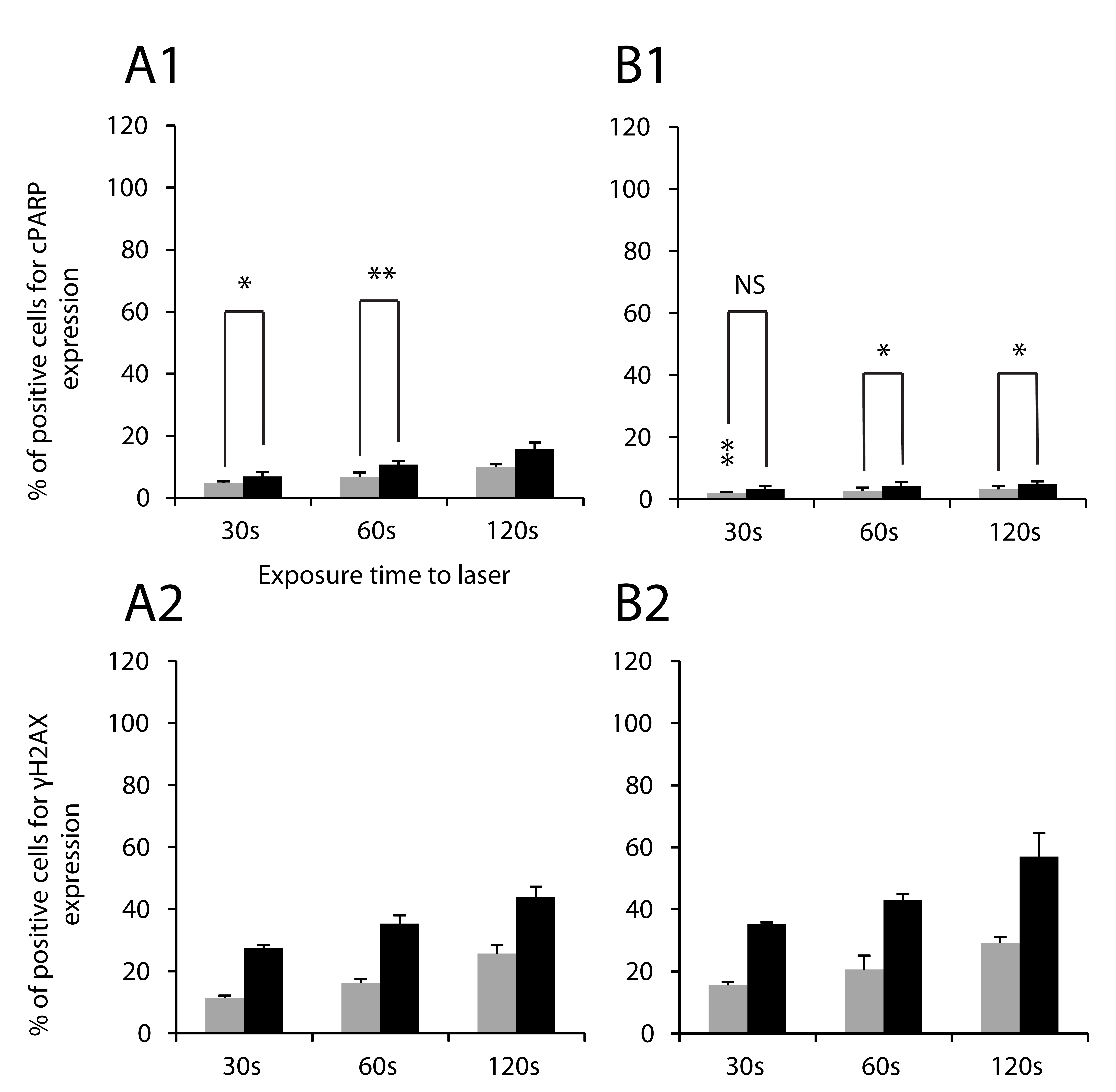}\caption{\textbf{Effect of laser intensity.}
Human lung-derived A549 (A) and HTB-182 (B) cancer cell lines exposed to vehicle (gray bars) or 100 $\mu$g/mL BFO HNPs (black bars), incubated 24 h and irradiated for 30, 60 or 120 s. Expression of cPARP (graph at the top) or $\gamma$H2AX (graph at the bottom) observed by IHC after further 24 h or 30 min of incubation, respectively. Results are expressed as \% ratio of positive cells. All comparisons between cells exposed to laser with and without BFO are significant (p<0.001) if not otherwise specified.}
\label{Fig:6}
\end{center}
\end{figure}

\begin{figure}
\begin{center}
\includegraphics[width=7 cm]{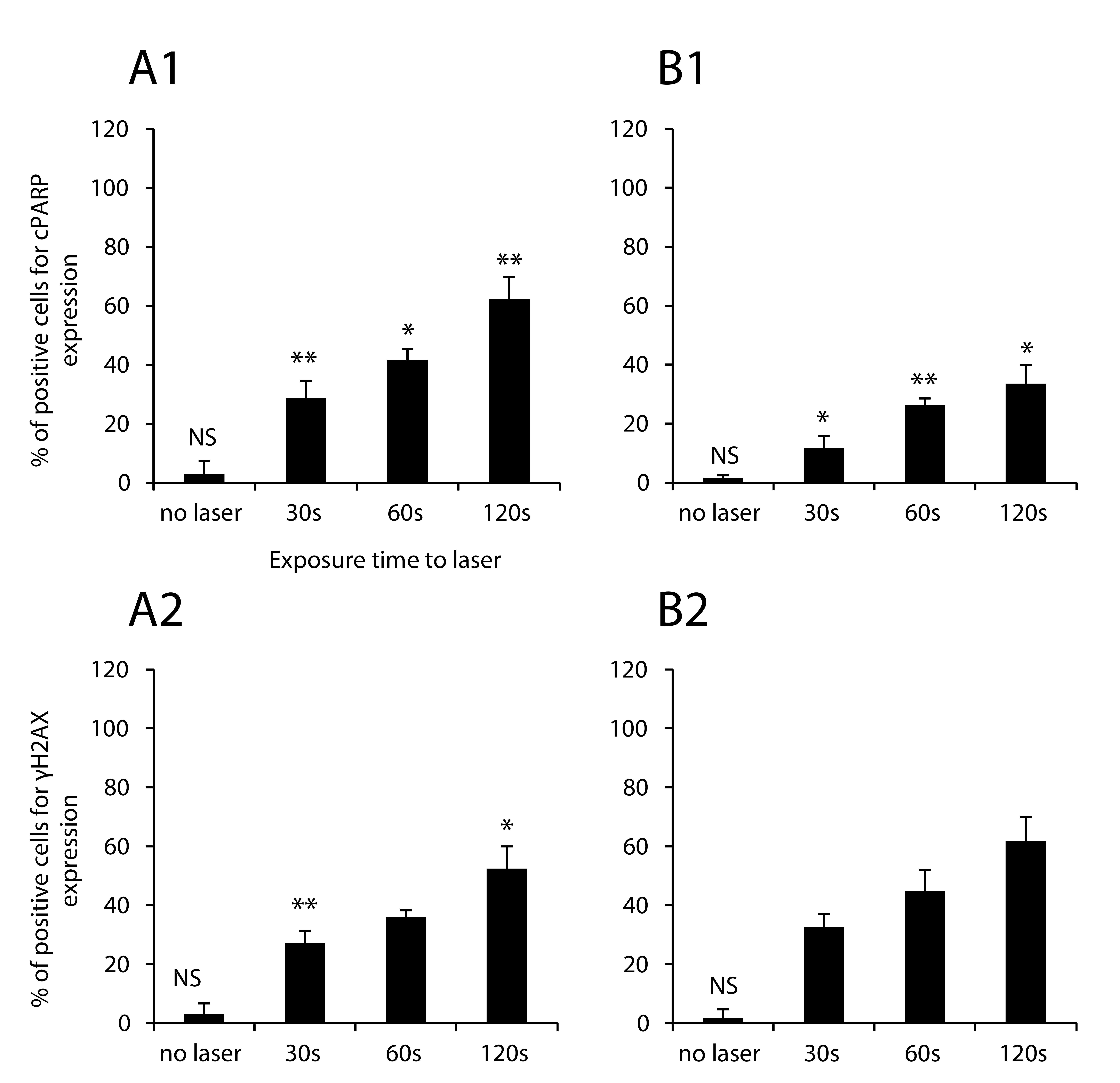}\caption{\textbf{Effect of HNPs concentration.}
Human lung-derived A549 (A) and HTB-182 (B) cancer cell lines exposed to 25 $\mu$g/mL HNPs, incubated 24 h and the irradiated for 30, 60, or 120s. Expression of cPARP (graph at the top) or $\gamma$H2AX (graph at the bottom) observed by IHC after further 24 h or 30 min of incubation, respectively. Results are expressed as \% ratio of positive cells. All comparisons to control or between cells exposed to laser with and without BFO are significant (p<0.001) if not otherwise specified.}
\label{Fig:7}
\end{center}
\end{figure}

To complete the characterization of the HNPs+visible irradiation sinergistic effects on cells viability,  we further investigated laser intensity and HNP concentration dependence. A549 and HTB-182 cell lines were exposed to 100 $\mu$g/mL HNPs or to vehicle and irradiated with the same laser average power focused onto a larger surface (400 $\mu$m diameter) corresponding to a neat fivefold intensity decrease with respect to the previous protocol. As reported in Fig. \ref{Fig:6}, a substantial decrease in the expression of both cPARP and $\gamma$H2AX was detected in the two lung-derived cancer cell lines. However, the difference between cells treated with BFO and with vehicle remained always significant, except for the expression of cPARP on HTB-182 cells after 30 s exposure, suggesting that BFO  generate DUV at lower irradiation intensity as well. The greatest difference between cells exposed to BFO and to vehicle was always observed for the longest exposure and the decrease on cPARP expression was greater compared to the expression of $\gamma$H2AX. For assessing the concentration dependence,  A549 and HTB-182 cells were treated with 25 $\mu$g/mL of HNPs and irradiated according to the high intensity protocol. Also in this case, as summarized in Fig. \ref{Fig:7}, cells  showed a reduced expression of cPARP, which remained anyway always significantly greater compared to that measured with vehicle. This reduction is more pronounced on HTB-182 cells. The number of positive cells for $\gamma$H2AX expression decreased in the two cell lines as well.

The results and controls presented altogether confirm that UV generation and related cytotoxic and genotoxic effects can be unambiguously ascribed to visible-light excited BFO HNPs. Major effects on cell viability (up to 100\% apoptosis) are observed in the irradiated areas. The difference on cPARP expression between  cells irradiated with lower laser intensity or exposed to lower particle concentration suggests that the synergistic effect between laser and HNPs is dominated by direct DNA photo-damages, but also implies other subsidiary mechanisms, such as a thermal effects and  cell membranes disruption.\cite{Aragane, Huang, Li} The  apoptotic cell fraction in samples exposed to 540 nm laser but not treated with HNPs can be ascribed to direct two-photon absorption by DNA, as previously observed.\cite{Konig, Tirlapur} Such a non HNP-specific interaction can be easily counteracted thanks to the fact that the cellular effects exerted by BFO HNPs are limited to the area of the laser spot, as highlighted in Fig. \ref{Fig:5}. If IR imaging is preliminary performed to precisely define the zone needing irradiation, treatment conserves its high specificity. It should be noted that treatment localization is expected to be greater in the proposed HNP-based approach, which is primarily based on direct UV absorption by DNA, than in photo-dynamic treatments involving upconverting and plasmonic NPs. In these two latter cases,  NPs-cell interaction is mediated by ROS, which are known to diffuse through tissues.\cite{Idris, Zhou}

In conclusion, we have presented and demonstrated  on human-derived cancer cell lines an original nano-theranostics approach based on the  nonlinear optical properties of  HNPs. The method proposed enables wavelength-selected imaging and  direct DUV photo-interaction with  nuclear DNA. The biocompatibility of BFO, a nanomaterial firstly applied here for biological applications,  screened for cytotoxicity and generation of oxidative stress, was found comparable to those of other HNPs or metal-based nanoparticles currently used in biomedical studies.\cite{Staedler, Barzilai, Frick, AshaRani}  It should be noted that all HNPs, possessing high nonlinear efficiency,\cite{Staedler}  can exert the effects described here,  which are therefore not unique to BFO.

DSBs DNA damages and induction of apoptosis are typical targets of photodynamic therapies, which normally involved the use of direct UV radiation (with poor tissue penetration and lack of specificity)  or chemical    photosensitizers.\cite{Offer, Soares}  To date, NPs-based strategies imply using of organic sensitizers (with some notable exceptions.\cite{Minai}) and are mediated by ROS generation.\cite{Ungun, Wang, Zhou, Barzilai} As for classical phototherapy, these approaches can generate major side effects due to the presence of toxic compounds and ROS which can diffuse to nearby tissues generating oxidative stress.\cite{Jomova, Scola} The  approach proposed, based on nonlinear optical response by HNPs and direct photo-interaction with nuclear DNA, might avoid side effects due to organic ligands and diffusion of toxic compounds, increasing selectivity and treatment localization. Moreover, the activation of the process is intrinsically limited to femtosecond-pulse excitation and cannot be obtained with any other artificial or natural light source, differently also from  the approaches based on sequential up-conversion of light frequency. This purely physical constrain greatly decreases the risk of unspecific treatment activation, in particular for surface lesions. Finally, and very notably, the proposed strategy allows to totally decouple diagnostic modality (IR imaging) from the therapeutic photo-dynamic action (visible irradiation), by simply tuning the excitation laser wavelength.

\subsection{Methods}

\paragraph{Multi-photon Imaging} The imaging set-up is based on a Nikon A1R-MP inverted microscope coupled with a Spectra-Physics Mai-Tai DeepSee tunable Ti:Sapphire oscillator. A Plan APO 40$\times$ WI N.A. 1.25 objective was used to focus the excitation laser and to epi-collect the nonlinearly excited signal (SH and membrane dye  fluorescence).  Four independent non-descanned detectors acquire in parallel the signal spectrally filtered by four tailored pairs of dichroic mirrors and interference filters (SH filter; 395$\pm$5.5, Chroma). Optimal pulse compression at the focal plane was adjusted by maximizing the SH signal of individual HNPs dispersed on a coverslip.

\paragraph{Visible laser irradiation}
For visible irradiation, we employed a two-stage non-collinear optical parametric amplifier (TOPAS White, Light Conversion) set at 540 nm. The output pulse characteristics are: 30 fs pulse duration, 15 mW average power, 1 KHz repetition rate. The sample was exposed for 30, 60 and 120 s using a laser spot size (measured by a high resolution beam profiler) of 170 $\mu$m or 400 $\mu$m diameter. During  irradiation, cells were kept at controlled temperature (37$^{o}$C), \ce{CO2} concentration (5\%) and humidity in a microscope incubator (Okolab UNO).

\paragraph{HNPs dispersion and characterization}
BFO NPs were provided by the German company FEE at high concentration  in ethanol. NPs were diluted 1:50 in 500 mL ethanol and decanted for 10 days. The supernatant was then taken, ethanol evaporated, and NPs re-suspended in distilled water. Successively, NPs  were dispersed by ultra-sonic bath for 24 h and quantified by Prussian blue assay. For this assay, 50 $\mu$L of  BFO solution were diluted in 50 $\mu$L HCl 6 M and 100 $\mu$L of 5\% potassium  hexacyanoferrate (Sigma-Aldrich) in PBS were added for 15 min. After incubation, the solution absorbance was measured at 690 nm in a multiwell-plate reader (Synergy HT, BioTek) and compared with the absorbance of a calibration curve with known BFO concentration. BFO NPs were finally diluted at 2 mg/mL in water. DLS and zeta-potential measurements were carried out with a Malvern NanoZ, yielding: zeta potential  -52.7 $\pm$ 3.5 mV,  mean hydrodynamic diameter 165.3 $\pm$ 24 nm.

\begin{acknowledgement}
This research has been conducted in the framework of European FP7 Research Project NAMDIATREAM (NMP4-LA-2010-246479, http://www.namdiatream.eu) and partially supported by the NCCR Molecular Ultrafast Science and Technology (NCCR MUST), a research instrument of the Swiss National Science Foundation (SNSF). The authors thank MER Dr Christine Wandrey for granting access to analytical equipment and S. Afonina and J. G\^ateau for assistance with  femtosecond laser equipment.
\end{acknowledgement}

\begin{suppinfo}
Additional experimental details are provided as supplementary information.

\end{suppinfo}

\bibliography{nanoUV}
\bibliographystyle{achemso}

\end{document}

\paragraph{Cell cultures}
Human lung-derived A549 and HTB-182 cancer cell line are available from ATCC (American Tissue Culture Collection). A549 were grown in Dulbecco's Modified Eagle Medium (DMEM) medium containing 4.5 g/L glucose, 10\% heat-inactivated fetal calf serum (FCS) and penicillin/streptomycin (PS) (all cell culture reagents were obtained from Invitrogen). HTB-182 were grown in complete Roswell Park Memorial Institute (RPMI) 1640 medium (Invitrogen) supplemented with 10\% FCS and PS.

\paragraph{Cell staining}
Cells were grown for 24 h and BFO HNPs at 50 $\mu$g/mL were added for further 24 h, then cell layers were fixed in 3\% formaldehyde in PBS, permeabilized 5 min in 0.1\% Triton X-100 (Sigma-aldrich) in PBS and exposed to 5 $\mu$g/mL of FM1-43FX fluorescent probe (Invitrogen, 1 mg/mL stock solution in DMSO) for 1 min on ice, washed with PBS and maintained in 4\% buffered formaldehyde at 4$^{o}$C until the acquisition of images.

\paragraph{Determination of cytotoxicity}
The cytotoxic effect of BFO HNPs at 100 $\mu$g/mL was determined after 5, 24 and 72 h incubation by the MTT assay as previously described.\cite{Staedler} Briefly, after incubation with the BFO or vehicle, MTT solution (3-(4,5-dimethyl-2-thiazoyl)-2,5-diphenyltetrazolium bromide (Sigma-Aldrich) 5 mg/mL in PBS) was added to the cells for 2 h. Then, the cell culture supernatants were removed, the cells layers were dissolved in 2-propanol/0.04N HCl and the absorbance at 540 nm was measured in a multi-well plate spectrophotometer (Synergy HT, BioTek). Experiments were conducted in triplicate, repeated twice and expressed as cell survival compared to cells exposed to vehicle. Means $\pm$ standard deviations were calculated.

\paragraph{Dithiothreitol assay}
BFO HNPs were diluted at 25 $\mu$g/mL or 100 $\mu$g/mL in nanopure water containing 250 $\mu$M DTT (Sigma-Aldrich) and incubated for 30 min at 37$^{\circ}$C, then 90 $\mu$L of Ellmann solution (5 mM 5,5'-Dithiobis-(2-nitroben- zoic acid) (DTNB), Sigma-Aldrich) were added and the absorbance was measured at 405 nm in a multiwell-plate reader (Synergy HT). DTT concentration was calculated by comparison with a standard DTT solution. Experiments were conducted in triplicate wells, repeated twice and converted as \% of consumed DTT. Means $\pm$ standard deviations were calculated.

\paragraph{Determination of reactive-oxygen species production}
Two different methods were used to determine ROS production in cells. In the former, ROS were detected by measuring the oxidation of DHE to ethidium. Cells were grown for 24 h and BFO at 25 $\mu$g/mL or 100 $\mu$g/mL were added for further 24 h. Then, cell layers were washed with PBS and 100 $\mu$M DHE (Sigma-Aldrich) in RPMI 1640 were added for 15 min at 37$^{\circ}$C. Cells layers were further washed and lysed with 0.1\% Triton X-100 (Sigma-Aldrich) in PBS. Ethidium production was determined in a fluorescent multiwell-plate reader (Synergy HT) at $\lambda_{ex}/\lambda_{em}$ = 485/580 nm. The latter method of ROS measurement involved the use of DCFH-DA.\cite{Frick} DCFH-DA is trapped inside cells as a sensitive cytosolic marker of oxidative stress, since its oxidation leads to the formation of the fluorescent dichlorofluorescein (DCF). Cells were grown and treated with BFO as previously described, then cells layers were washed with PBS and exposed to 20 $\mu$M DCFH-DA in Hank's buffer solution (HBSS) (both from Invitrogen) for 40 min at 37$^{o}$C. After incubation, the cell layers were further washed and fluorescence of DCF was measured at $\lambda_{ex}/\lambda_{em}$ = 485/527 nm. Experiments were conducted in triplicate, repeated twice and expressed as fold-increase compared to cells exposed to vehicle. Means + standard deviations were calculated.

\paragraph{Immunohistochemistry}
Fixed cell layers were washed twice with PBS, permeabilized with cold methanol (-20$^{\circ}$C) for 5 min, washed with PBS and incubated 10 min at RT in 3\% H2O2 in methanol. Then, cells were washed and incubated for 1.5 h at RT with anti-phospho-histone H2AX (Ser139) or anti-cleaved PARP antibodies (both from Cell Signaling) diluted 1:250 in Dako REALTM Antibody Diluent (Dako). After incubation, cell layers were washed twice with PBS and incubated with undiluted anti-rabbit HRP-conjugate antibody (EnVision+ Sytem, Dako) for 30 min at RT. Cells were further washed and HRP activity was revealed using the DAB+ CHROMOGEN system from Dako, according to the supplier instructions. Finally, cells were counterstained with hematoxylin and images were taken using a microscope (DM IL LED from Leica) equipped with a digital camera (ICC50HD, Leica). Three pictures per treatment were taken and positive cells in a surface of 0.3 mm$^2$ around the laser-spot were counted using the ImageJ software. Experiments were conducted in triplicate wells, repeated twice and expressed as \% of positive cells per picture.

\paragraph{Statistical analysis}
Data were compared using a homoscedastic, two-tailed distributed Student's t-test. Details about comparisons are specified in the caption of each figure. Significance is expressed as: NS $p>0.05$; * $p<0.05$; ** $p<0.01$; *** $p<0.001$.